\documentclass[useAMS,usenatbib]{mn2e}
\usepackage{amsmath}
\usepackage{amssymb}
\usepackage{graphicx}

\newcommand{\Me}{{\rm M}_{\oplus}}

\title[On the corotation torque for low-mass eccentric planets]
{On the corotation torque for low-mass eccentric planets}
\author[S. M. Fendyke and R.P. Nelson]{Stephen M. Fendyke\thanks{E-mail:
S.M.Fendyke@qmul.ac.uk} and Richard P. Nelson\\
Astronomy Unit, School of Physics and Astronomy, Queen Mary, University of London, United Kingdom}
\begin{document}

\date{Accepted 2013 September 30.  Received 2013 September 30; in original form 2013 July 12}

\pagerange{\pageref{firstpage}--\pageref{lastpage}} \pubyear{2013}

\maketitle

\label{firstpage}

\begin{abstract}

We present the results of high resolution 2D simulations of low mass planets 
on fixed eccentric orbits embedded in protoplanetary discs. The aim of this study 
is to determine how the strength of the sustained, non-linear corotation torque 
experienced by embedded planets varies as a function of orbital eccentricity, disc 
parameters, and planetary mass. In agreement with previous work we find that the corotation torque diminishes
as orbital eccentricity, $e$, increases. Analysis of the time-averaged streamlines in 
the disc demonstrates that the width of the horseshoe region narrows as the
eccentricity increases, and we suggest that this narrowing largely explains the observed 
decrease in the corotation torque. We employ three distinct methods for estimating the strength of the unsaturated 
corotation torque from our simulations, and provide an empirical fit to these 
results. We find that a simple 
model where the corotation torque, $\Gamma_{\rm C}$, decreases exponentially 
with increasing eccentricity (i.e. $\Gamma_{\rm C} \propto \exp{(-e/e_{\rm f})}$) 
provides a good global fit to the data with an e-folding eccentricity, $e_{\rm f}$, that scales 
linearly with the disc scale height at the planet location. We confirm that this 
model provides a good fit for planet masses of 5 and 10 $\Me$ in our simulations. The formation of planetary systems is likely to involve significant planet-planet
interactions that will excite eccentric orbits, and this is likely to influence 
disc-driven planetary migration through modification of the corotation torque. Our 
results suggest that high fidelity models of planetary formation should account
for these effects.

\end{abstract}


\section{Introduction}

The current dataset describing the observed population of extrasolar planets
displays a broad diversity in physical and orbital properties. Inspection of
the currently confirmed exoplanets\footnote{see http://exoplanets.org} \citep{exoplanetorg}
reveals the existence of numerous short-period massive planets (`hot-Jupiters'),
and multiple planet systems composed of compact, short-period bodies of low and intermediate mass
(super-Earths and Neptune-like planets). Examples of these latter systems include
Kepler-11 with six planets \citep{lissauer2011closely}, Kepler-20 with five 
\citep{gautier2012kepler}, Kepler-62 with five \citep{borucki2013kepler} and 
HD 10180 with up to seven detected by the HARPS spectrograph \citep{lovis2011harps}.
Numerous other multi-planet systems have also been reported in the literature
\citep{endl2012revisiting,fabrycky2012transit,lissauer2012almost,gl581}.
Taken as a whole, these planetary systems are likely to contain substantial
mass in heavy elements, such that their existence is difficult to 
explain using {\it in situ} formation scenarios because most disc models
contain insufficient inventories of solid material at small radii 
\citep[e.g.][]{1981PThPS..70...35H,1977Ap&SS..51..153W}. Large scale migration, 
possibly coupled with continuing mass growth, would appear to provide the most 
compelling explanation for many of these systems, although N-body models coupled 
with disc-driven migration have so far not managed to reproduce short-period 
multi-planet systems that are particularly similar to those observed 
\citep[e.g.][]{2010MNRAS.401.1691M}.

While planet-planet gravitational scattering coupled with tidal interaction
with the central star may explain some short-period planets, the compact, 
low mutual-inclination, short-period systems such as Kepler-11 appear to be best 
explained through gas disc-driven migration. Low mass planets whose
Hill radii are smaller than the local scale height (such that they
do not carve out deep, tidally-truncated gaps) experience type I migration, 
driven by a combination of Lindblad and corotation torques 
\citep{1980ApJ...241..425G,1997Icar..126..261W,2002ApJ...565.1257T}. Particular interest has 
focussed on the role of corotation torques since it was first realised that 
they may counterbalance the rapid inward migration driven by Lindblad torques.
In particular, strong positive gradients in disc surface density can cause the 
corotation torque to stall migration due to the associated gradient %
in vortensity \citep{2006ApJ...642..478M}, 
and a negative entropy gradient may also cause migration to stall 
\citep{paardekooper2006halting,baruteau2008corotation,paardekooper2008disc}.

Strong corotation torques (also known as ``horseshoe drag''\citep{ward1991}) arise through
interaction between the planet and gas that executes horseshoe orbits in a disc 
with a radial gradient in vortensity and/or entropy. Given that horseshoe streamlines 
are a non-linear phenomenon, horseshoe drag is also referred to as the non-linear 
corotation torque \citep{paardekooper2008disc}. The vortensity-related corotation torque 
is prone to saturation in the absence of viscosity, which maintains the vortensity 
gradient across the horseshoe region against the tendency of orbital phase mixing 
there to flatten it out. Similarly, thermal diffusion or cooling is required to
maintain the entropy-related corotation torque against saturation.
Torque formulae have been derived that allow the steady-state corotation torque
to be calculated for a broad range of disc and planet parameters
\citep{paardekooper2009torque,paardekooper2011torque,2010ApJ...723.1393M}.

In addition to saturation in the absence of viscous or thermal diffusion,
the corotation torque has been shown to diminish if the planet orbit
becomes eccentric \citep{bitschkley2010}. At present the physical reason for this 
decrease is not clear, and as yet there has not been an extensive analysis
of how the dependence of the corotation torque on eccentricity scales with variations 
in disc and planet parameters. Given that planet-planet interactions during planet
formation and migration lead inevitably to eccentricity excitation 
\citep[e.g.][]{cresswell2006evolution}, further exploration of these issues 
is important in order to fully understand the role of migration in planetary formation.
Using simple N-body simulations of planetary accretion coupled with prescriptions
for type I migration torques obtained from \citet{paardekooper2011torque},
\citet{hellarynelson2012} examined the possible influence of eccentricity excitation
on the oligarchic growth of planets, and concluded that the ability of horseshoe
drag to prevent rapid inward migration of growing planets is diminished strongly
when the associated quenching of the corotation torque is accounted for.
Further examination of this is clearly required to test the assumptions
of how the torque scales with eccentricity adopted in this latter study.

In this paper, we present results from 2D hydrodynamic simulations of eccentric planets of
different mass embedded in protoplanetary discs with differing effective vertical scale
heights. Particular challenges faced when analysing the results include the tendency for
moderate gaps and vortices to form in low viscosity discs with relatively small vertical 
scale heights (i.e. $H/r \lesssim 0.05$). 
To overcome these problems three different methods for estimating the unsaturated corotation 
torque were employed. As expected from the earlier simulations of 
\citet{bitschkley2010}, we observe that the corotation
torque decreases as the planet eccentricity increases. We provide an empirically
derived analytic fit formula for our simulation results which shows that the
corotation torque decreases exponentially with orbital eccentricity, with
the e-folding eccentricity scaling linearly with the local disc scale height.

The paper is organised as follows: 
In section 2 we discuss our methodology and simulation setup. 
In section 3 we present the simulation results, and in section 4 we discuss and interpret 
these results. We draw our conclusions in section 5.

\section{Methodology}

\subsection{Disc model}

To compute the disc model, we use a modified version of the magnetohydrodynamic code 
NIRVANA \citep{ziegler1998nirvana}, which is based on the ZEUS algorithm 
\citep{stonenorman1992}, to solve the vertically integrated hydrodynamic equations
in polar coordinates ($R$, $\phi$):

\begin{align}
\partial_t \sigma + \nabla \cdot (\sigma \vec{v}) & = 0 \\ 
\partial_t (\sigma \vec{v}) + \nabla \cdot [\sigma \vec{v} \vec{v}] & = -\nabla P - \sigma \nabla \Phi \\ 
\partial_t e + \nabla \cdot (e \vec{v}) & = -P \nabla \cdot \vec{v} + {\cal Q} - \Lambda
\end{align}
where $e$ is the internal energy density, ${\cal Q}$ and $\Lambda$ are heating and 
cooling terms, $v$ is velocity, $P$ is pressure,  $\sigma$ is the surface density and 
$\Phi$ is the combined gravitational potential of the central star and the planet.
We use an ideal gas equation of state to close the system 
of equations: $P=(\gamma-1) e$ with the adiabatic exponent $\gamma=1.4$.

We simulate an embedded planet in a non self-gravitating two-dimensional disc 
with inner and outer radii located at 0.5 and 1.8 au, respectively. Reflecting boundary 
conditions are employed at the radial boundaries, in conjunction with damping zones that
minimise wave reflection using the scheme presented in \citet{val2006comparative}. 
We use a resolution of either 1020 or 1024 cells in radius and 2048
in azimuth. The simulations are computed in a frame corotating with the guiding 
centre of the planet, such that the eccentricity is manifest as epicyclic motion 
around this guiding centre. 
The gravitational force of the planet acting on the disc is softened
using a softening parameter, $b=0.4h$.  The disc density profile takes the form 
$\sigma = \sigma_0 r^{-\alpha}$ with $\alpha=0.5$, and the temperature profile is
$T = T_0 r^{-\beta}$ with $\beta=2.0$. Our disc mass is normalised by
$\sigma_0 = 1.35 \times 10^{-3}$.

We consider planets with mass ratios $q = 1.5 \times 10^{-5}$ and $3 \times 10^{-5}$ 
(equivalent to 5 and 10 Earth mass planets orbiting a solar-type star). The orbit
remains fixed, and is integrated using a fifth order Runge-Kutta scheme, while 
the torque experienced by the planet is recorded as a time series.

As described in the introduction, preventing saturation of the entropy related corotation 
torque requires thermal diffusion of the gas so that the entropy gradient across
the horseshoe region is maintained. Rather than including a computationally expensive 
full-blown radiative transfer model, we have implemented a simple Newtonian cooling
scheme that constantly forces the entropy in the disc back toward its initial value
on a specified time scale, $\tau_{\rm ent}$. The value of $\tau_{ent}$ is chosen
through experiment to optimally unsaturate the corotation torque.
Given that $P/\sigma^\gamma$ is a constant along each adiabat and is therefore a function of entropy, $s$, we define the function
\begin{equation}
K(s) = \frac{P}{\sigma^\gamma},
\end{equation}
and iterate it at each timestep according to %
\begin{equation}
K_{i+1}(s) = K_i(s) - (K_i(s)-K(s_0))\frac{\Delta t}{\tau_{ent}},
\end{equation}
where $K(s_0)$ is the initial value of the entropy function and $\Delta t$ is the 
time-step size. From this, the internal energy density is recalculated using the 
ideal gas law
\begin{equation}
e = K(s) \frac{\sigma^\gamma}{\gamma-1}.
\end{equation}

\subsection{Simulations}
\label{sec:sims}
\begin{figure}
 \includegraphics[width=0.5\textwidth]{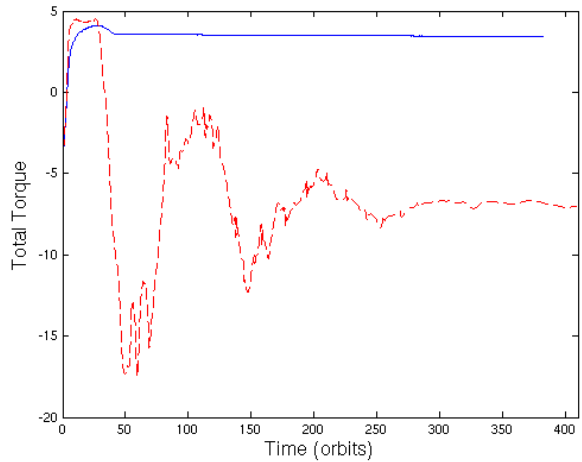}
 \caption{Torque time series for a disc with optimally unsaturated corotation torque (blue, solid line) and saturated torque (red, dotted line) respectively. Time is given in orbits and torque is shown in units of $\Gamma_0 / \gamma$. In both simulations, the disc thickness is $H/r = 0.05$ and the planet is fixed on a circular orbit.}
 \label{fig::optimised}
\end{figure}

We computed disc models with aspect ratios $h \equiv H/r=0.03$, 0.05, 0.07 and 0.1.
For each value of $h$, eccentricity values in the interval $0 \le e \le 0.3$ were 
considered. For each combination of $h$ and $e$, we performed two separate simulations:

\noindent
(A) An adiabatic disc with viscosity at the lowest value consistent with obtaining a
well-behaved time series for the measured torque on the planet. We give these values in table~\ref{table:optim_params}. In the absence of this
small viscosity we find that torque time series may become difficult to interpret
due to strong time-dependencies introduced through the development of small
vortices that form near the separatrices between circulating and librating material.
A low viscosity adiabatic disc allows saturation of both the vortensity and entropy related 
contributions to the corotation torque as material in the horseshoe region becomes phase 
mixed and the vortensity and entropy gradients disappear. A torque time series from 
one of these simulations when the planet is on a circular orbit is shown in 
figure~\ref{fig::optimised}.

\noindent
(B) A disc with viscosity and thermal diffusion set to values selected by successive trials 
to optimally unsaturate the corotation torque. We give the optimal values in 
table~\ref{table:optim_params}. The torque time series from such a simulation with a circular 
orbit is shown in figure~\ref{fig::optimised}. We note that all torques plotted in
this paper are normalised by the quantity $\Gamma_0/\gamma$ where
\begin{equation}
\label{eq:gam0}
\Gamma_0 = \left(\frac{q}{h}\right)^2 \sigma_{\rm p} r_{\rm p}^4 \Omega_{\rm p}^2,
\end{equation}
and $q$ is the planet-star mass ratio $m_{\rm p}/M_*$, $r_{\rm p}$ is the planet
orbital radius, and $\Omega$ is the Keplerian angular velocity. A subscript `p' denotes
evaluation at the planet location.

For all simulations, as the non-linear horseshoe drag is due to material undergoing 
horseshoe orbits, and the timescale for even the shortest horseshoe orbit is significantly 
longer than the planetary orbital period, we continually construct and record 
time-averaged density and velocity fields from the disc for further analysis. We
also maintain a record of the contribution to the torque exerted on the planet
by the disc as a function of radius in the disc.

\begin{table}
\caption{Optimised entropy relaxation timescale, $\tau_{ent}$, optimised kinematic viscosity, 
$\nu_{opt}$, and minimum kinematic viscosity, $\nu_{min}$, that allows for a quasi-steady
flow for a planet on a circular orbit in a disc of various thicknesses. These values are 
obtained through numerical experimentation.}
\begin{center}
    \begin{tabular}{ | l | c | c | c | }
    \hline
    $H/r$ & $\tau_{ent}$ (orbits) & $\nu_{opt}$ & $\nu_{min}$\\ \hline
    0.03 & 7 & $8 \times 10^{-6}$ & $2.5 \times 10^{-7}$ \\ 
    0.05 & 11 & $5 \times 10^{-6}$ & $1 \times 10^{-7}$ \\ 
    0.07 & 13 & $4 \times 10^{-6}$ & $1 \times 10^{-7}$ \\ 
    0.10 & 14 & $1.5 \times 10^{-6}$ & $1 \times 10^{-8}$ \\ 
    \hline
    \end{tabular}
\end{center}
\label{table:optim_params}
\end{table}

For simulations in set A, it is assumed that after the total torque has reached 
a steady-state, the corotation torque has saturated and only the Lindblad torque remains. 
For those in set B, we expect the steady-state total torque to approach the value 
predicted by the formulae of \citet{paardekooper2009torque} for unsaturated
corotation torques in the circular orbit case.
Using these two sets of simulations, we estimate the corotation torque for each
different eccentricity using three methods:

\noindent
(i) By taking the difference between the long-term, steady-state, time-averaged torques 
obtained in the
corresponding simulations from sets A and B. This method assumes that the Lindblad
torque is the same in each simulation, such that the torque difference measures
the unsaturated corotation torque obtained in the set B simulation directly.

\noindent
(ii) By using time-averaged velocity fields to determine the extent of the region in which 
disc material is undergoing horseshoe orbits. We aggregate the torque contribution from 
this region by using time-averages of the torque versus radius data that we accumulate
during the simulations.

\noindent
(iii) By measuring the initial peak in the torque time series from simulations in set A
associated with the initial growth of the corotation torque prior to its long-term
saturation. This is the method most similar to previous measurements of the corotation 
torque \citep{paardekooper2009torque}.  The peak is then compared with the long-term
steady torque which is assumed to comprise the Lindblad torque only due to saturation
of the corotation torque.

\begin{figure}
 \includegraphics[width=0.5\textwidth]{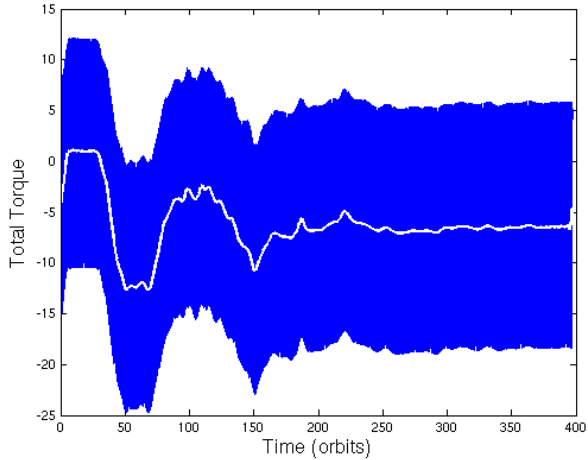}
 \caption{A torque time series for a planet with eccentricity, $e=0.02$ in a disc with $h=0.05$ (blue). 
Superimposed is the same time series with the high frequency oscillations in torque due to the planet's 
epicyclic motion filtered out (white).}
 \label{fig::fourierfilter}
\end{figure}

\subsection{Torque filtering}

The torque time series for an eccentric planet consists of an average value, due to the 
motion of the planet's guiding centre, and a contribution due to the epicyclic motion
that varies quasi-periodically on the planet orbital period. At high eccentricities, 
the contribution from the epicyclic motion dominates and obscures the averaged value 
that we require. As this motion occurs on a much more rapid timescale than that of material 
within the corotation region, we use a Fourier transform filter to remove oscillations 
occurring on timescales more rapid than a few orbital periods. The result
of employing this procedure is shown in figure \ref{fig::fourierfilter}.

\section{Results}

In this section we present the results from the simulations. We show 
time-averaged surface density fields for a sample of the runs in which sustained, unsaturated
corotation torques were obtained, and the magnitudes of corotation torques estimated using
the three methods described in section~\ref{sec:sims}. For each set of corotation torque results 
we obtain a simple analytical fit that describes the variation of corotation torque with
eccentricity.

\begin{figure*}
\begin{minipage}{126mm}
 \includegraphics[width=\textwidth]{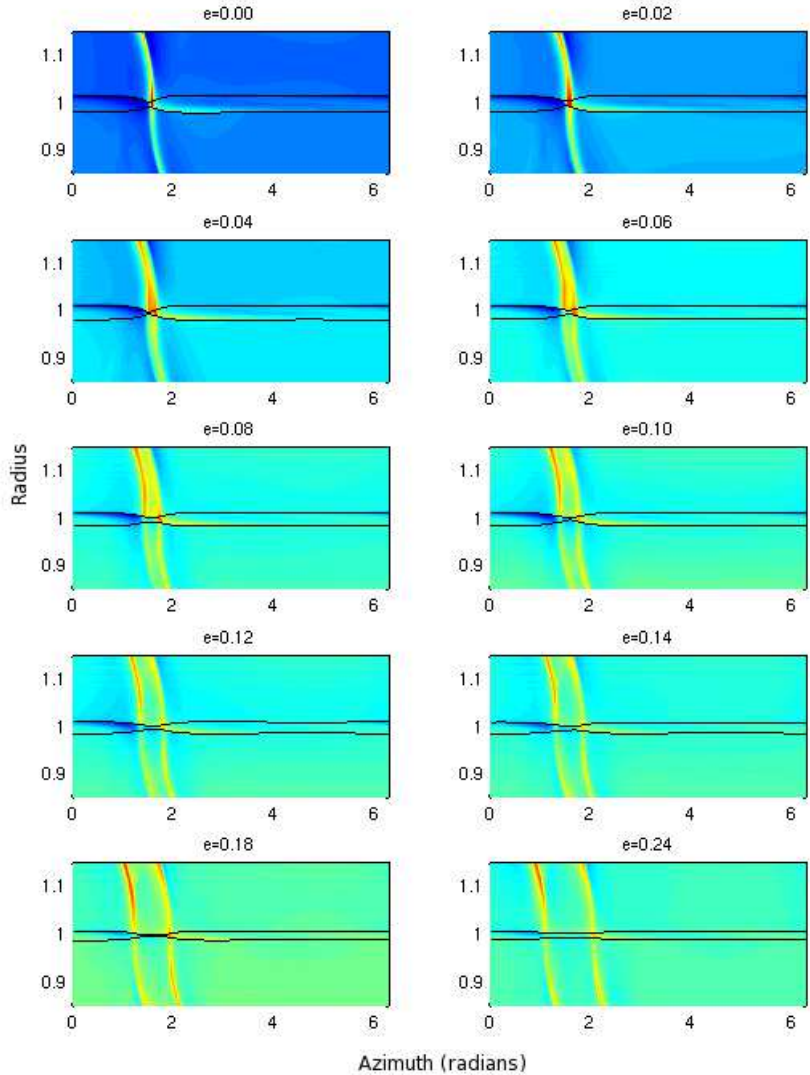}
 \caption{The time-averaged fractional perturbation of the initial disc surface density, with the last circulating 
streamline superimposed. Fields are plotted between 0.85 and 1.1 orbital radii of the planet and for the entire 
azimuth of the disc, for eccentricities between 0 and 0.24. The colours used are not to the same scale. The disc 
used to generate these fields has aspect ratio, $h = 0.07$. Note (i) the narrowed horseshoe region for increased 
eccentricities, which we associate with decreased corotation torque; (ii) the `splitting' of the distinctive 
spiral density wave into two strands at increased eccentricity; and (iii) the presence of a sustained density 
perturbation within the corotation region (between the last circulating streamlines).}
 \label{fig::densityfields}
 \end{minipage}
\end{figure*}

In figure \ref{fig::densityfields} we show time-averaged steady-state surface 
density fields for the $h=0.07$ 
discs with optimised, sustained corotation torques. Using the corresponding 
time-averaged velocity fields, 
we locate and superimpose the circulating streamline that sits closest to 
the horseshoe region (interior and exterior to the planet) where the gas is seen 
to librate rather than circulate. As such, this streamline 
acts as the boundary that separates the circulating and librating regions. 
Inspection of figure~\ref{fig::densityfields} shows clearly the tendency for the width of the
horseshoe region to decrease as the eccentricity increases, with the horseshoe streamline u-turns broadly confined to the region outside the path of the planet's epicyclic motion. Furthermore, each panel shows the presence of a 
positive surface density perturbation within the horseshoe region that sits just ahead of the planet, and a 
negative perturbation that sits just behind it (in the inertial frame, the sense of motion in this figure would
be from left to right). As has been discussed in previous work \citep[e.g.][]{baruteau2008corotation}, these perturbations 
arise from the advection of fluid elements on horseshoe orbits that almost conserve their entropy (and vortensity) 
around the horseshoe u-turn (in the absence of viscosity, thermal relaxation or shocks these quantities should 
be conserved). Maintenance of local pressure equilibrium causes regions that receive low entropy material from 
the outer disc to contract. Regions behind the planet that receive high entropy material expand. The resulting 
surface density perturbations lead to the observed positive corotation torque, as shown earlier in figure~1, for 
example. This perturbation is present in all simulations presented in figure~\ref{fig::densityfields},
but diminishes as the eccentricity increases and the width of the horseshoe region decreases because the
advected entropy introduces a reduced pressure perturbation.

Each of the panels in figure~\ref{fig::densityfields} shows the characteristic spiral density wave, but
as the eccentricity increases the single wave that is present interior and exterior to the planet splits into
two well-defined wake-like structures. This arises because the epicyclic motion of the planet around its 
guiding centre causes the planet to travel more slowly than local disc material when at apocentre, and
faster than local material when at pericentre. This leads to periodic excitation of inward and outward
propagating wakes at these two phases of the orbit, as described for example in 
\citet{kleynelson2012}\footnote{An animation associated with this review article, showing the influence of a 
30 $\Me$ planet on an eccentric orbit with $e=0.1$ embedded in a disc with $H/r=0.05$, may be seen at 
http://www.youtube.com/watch?v=65nqq9sEZdM}. 
The relative motion between gas and planet at apo- and pericentre also leads to a reversal of the normally 
negative Lindblad torque when $e \ge h$. This is because orbiting gas is gravitationally focussed to a 
region that leads the planet at apocentre, creating a positive density perturbation in front of the planet
that exerts a positive torque. This provides the dominant contribution to the orbit-averaged torque because 
the planet spends most of its time at apocentre. This reversal of the Lindblad torque for $e > h$ was
first reported by \citet{papaloizou2002orbital}, who presented torque calculations based on summing
contributions from numerous eccentric Lindblad resonances. 

We now present results on how the corotation torque varies with eccentricity and disc parameters,
showing the results obtained with each of the methods used to estimate the corotation torque.
We begin by presenting results for planet mass 5 $\Me$ and disc models with $h=0.03$, 0.05 and 0.07,
followed by a model with 5 $\Me$ and $h=0.1$. Finally, we present a model with planet mass 10 $\Me$
and $h=0.1$.

\subsection{Method (i): Comparing torque time series}

\begin{figure}
 \includegraphics[width=0.5\textwidth]{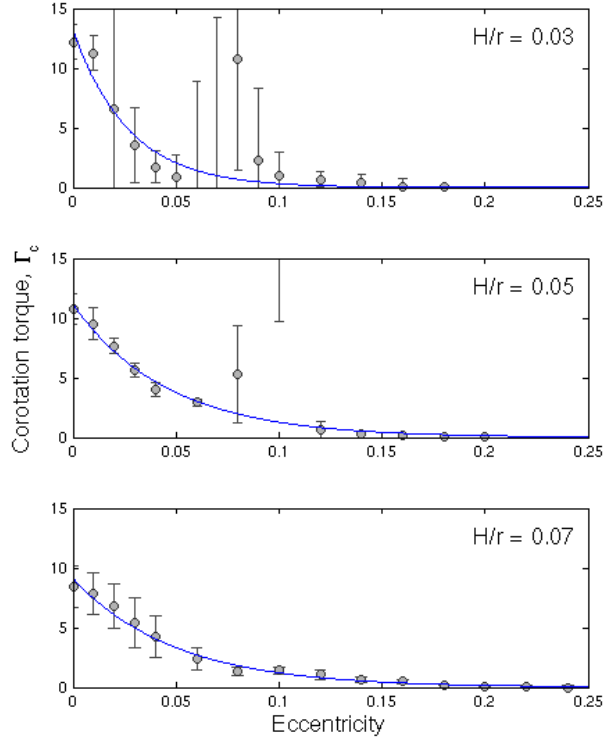}
 \caption{We plot the corotation torque measured by method (i); by taking the difference in steady state total torque 
between simulations with a saturated and optimally unsaturated corotation torque (sets B and A respectively). 
High-frequency oscillation in the torque due to the epicyclic motion of the planet has been filtered out, and the 
error bars shown are three times the combined standard deviations of the mean steady state torques. Results are 
given for three disc aspect ratios and fits are of the form $\Gamma_c = \Gamma_{c,e=0}\exp\left(-\frac{e}{e_{\rm f}}\right)$. 
Corotation torque is given in units of $\Gamma_0/\gamma$.}
 \label{fig::method_i_and_fit}
\end{figure}

When using this method to estimate the steady corotation torque as a function of eccentricity,
we first filter out high frequency oscillations in the torque time-series
due to the epicyclic motion of the planet, and then take the difference between the 
results of simulations in sets A and B. 
In principle, the long-term steady torques in set A converge to pure Lindblad torques,
and those in set B consist of the Lindblad plus sustained corotation torque, so we 
take the difference and use this as a measure of the corotation torque. 
The results obtained using this method are shown in figure~\ref{fig::method_i_and_fit},
where we plot the estimated corotation torque value versus eccentricity.
As there is residual time variation in the torque at the end of the simulations,
we plot the mean torque (averaged between 240 and 350 orbits),
and error bars showing three standard deviations about the mean. 

The simulations from set B with thermal relaxation and viscosity produce smooth,
well-behaved results that tend toward a well-defined steady state after sufficient
run-time.
There are a number of issues, however, affecting some low-viscosity
disc models from set A that
combine to make it difficult to obtain accurate estimates of corotation torques.
They involve restructuring of the disc in some fashion.
First, we find that the thinner, low viscosity disc models 
develop moderate gaps due to tidal torques from the planet. 
These have depth $\sim 10$-20 \% of the background surface density.
Although these models do not satisfy the usual gap formation requirement that the planet Hill 
sphere size exceeds the vertical scale height 
\citep[see e.g.][]{1993prpl.conf..749L,crida2006width},  non-linear damping of the spiral 
waves deposits angular momentum in the disc near the planet and can cause a 
moderate annular dip to develop in the local surface density profile. 
This effect has been predicted analytically by \citet{2002ApJ...572..566R} and 
observed in simulations by \citet{muto2010}. The gap impacts 
on the estimate of the Lindblad torque in these cases, and therefore affects the 
corotation torque estimate because a similar gap does not develop in the
corresponding viscous disc model. 

The second issue is that for eccentricities
$e \ge 0.06$ in the $h=0.03$ runs, a large scale discrete vortex forms very close
to the corotation radius of the planet guiding centre. This moves very slowly
relative to the planet, but exerts a time-varying torque on it that is 
very difficult to average out because of the long run times that would be required.
This clearly has an effect on our ability to measure the corotation
torque using method (i). Finally, for some runs with $h=0.03$ and 0.05, and for intermediate values
of the eccentricity (i.e. $0.07 \lesssim e \lesssim 0.1$), we observe the development of discrete
structures in the simulations.
When the time averaged surface density is rendered using contour plots
similar to those shown in figure~1 we observe these structures to sit
in or at the inner edge of the horseshoe region close to the mean location of the 
planet. They do not appear to be vortices, but instead seem to be features
related to the high density structure that forms behind the planet at pericentre.
As can be seen in the middle panel of figure~4, for example, this is an issue that
affects the torque for the $e=0.08$ and 0.1 cases for this particular value of $h$. So far 
we have been unable to determine why only these specific runs give rise to this phenomenon.

\begin{figure}
 \includegraphics[width=0.5\textwidth]{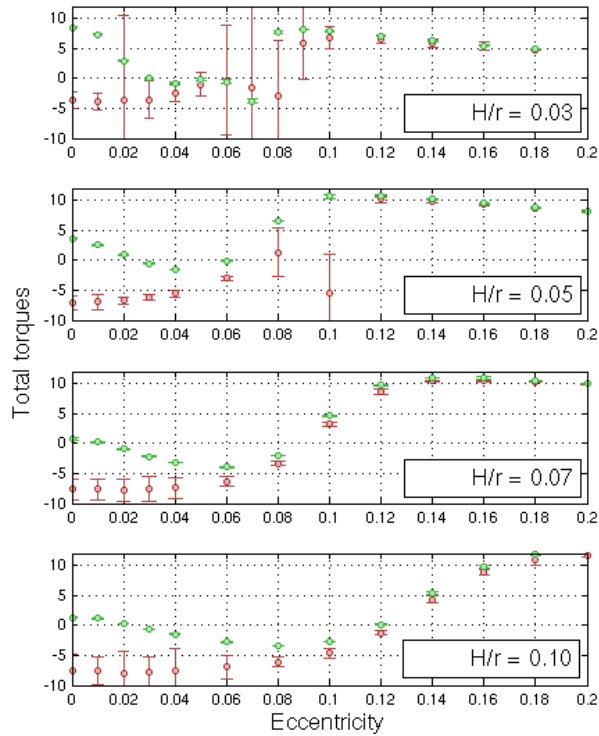}
 \caption{Total torque as a function of eccentricity for simulations from 
set A (green) and set B (red). Note the transition in torque from negative to 
positive values at around $1.1h$.}
 \label{fig::total_torque}
\end{figure}

In figure \ref{fig::total_torque}, we show the total torques as a function
of eccentricity from all our simulations with planet mass 5 $\Me$ and $h=0.03$, 
0.05 and 0.07. These are the torques used by method (i) to calculate the corotation 
torque, and are particularly of note because they display the total torque 
(including the unsaturated corotation torque) and the Lindblad torque.  
In agreement with \citet{papaloizou2002orbital}, the Lindblad torque changes sign 
from negative to positive values for $e > 1.1h$, and we see that at large values 
of $e$ the two sets of torque values essentially coincide as the corotation 
contribution diminishes.

\subsection{Method (ii): Streamline-defined horseshoe region}
\begin{figure}
 \includegraphics[width=0.5\textwidth]{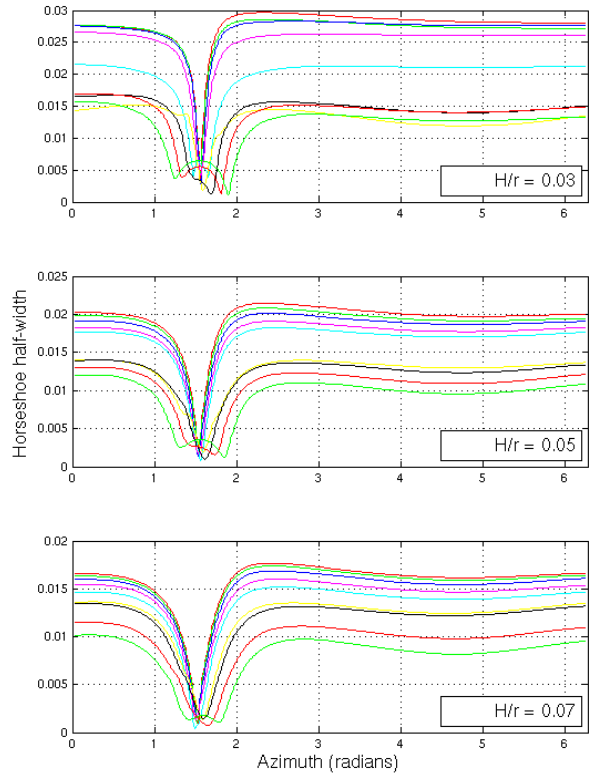}
 \caption{For three different disc aspect ratios, and eccentricities in the range 0 to 0.3,
 we plot the horseshoe half-width,
defined here as half the separation between the inner and outer last circulating streamlines, 
as a function of azimuth. As the eccentricity increases, the overall horseshoe width decreases and a 
``double-cusp'' reflects the planet's epicyclic motion for larger values of $e$. These 
values were calculated using discs with optimally unsaturated steady-state corotation torques. }
 \label{fig::hshoewidth}
\end{figure}

This method estimates the corotation torque by defining a region of the disc
to be the horseshoe region through inspection of fluid streamlines obtained from
time averages of the disc velocity field taken over many orbits of the planet.
A starting location is chosen in the disc from which we construct fluid streamlines by integrating the
averaged velocity field. Bilinear interpolation is used to define the local velocity away from the centres 
of grid cells. By performing this integration for a large number of closely separated initial locations, 
we are able to precisely locate the region in which material undergoes horseshoe turns on average.
In our parlance, the location between librating and circulating material is delineated
by the `last circulating streamline'. There is one interior and exterior to the planet's
semi-major axis. In figure \ref{fig::hshoewidth} we plot half the distance between the inner 
and outer last circulating streamlines as a function of azimuth for all models drawn from
set B that we are considering in this section ($m_{\rm p}=5\; \Me$ and $h=0.03$, 0.05, 0.07). 
Each line corresponds to a simulation with different planet eccentricity, and we note the clear 
trend for the horseshoe region to narrow as $e$ increases. 
The half-width of the corotation region is smallest for the most eccentric planets in the 
thickest disc, where $x_s \sim 0.01$, and we note that this is resolved by $\sim 10$ cells 
in our simulations. We remind the reader that
for the $h=0.07$ disc, these streamlines are also shown in figure \ref{fig::densityfields}.
We have noted previously that they narrow for increasing eccentricity, and that
the density perturbations associated with the corotation torque are contained within
the defined boundaries of the corotation region.

\begin{figure}
 \includegraphics[width=0.5\textwidth]{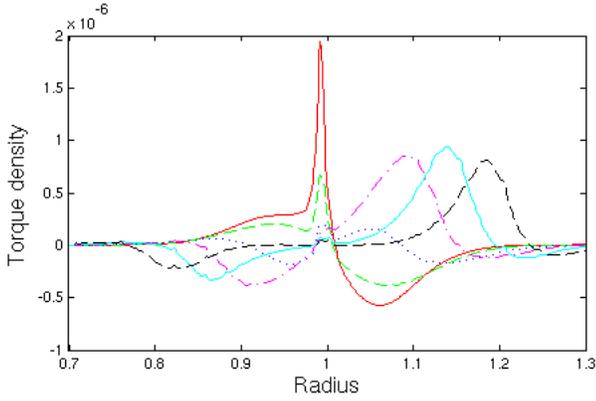}
 \caption{Torque as a function of radius for a disc with $h = 0.07$ and a sustained 
corotation torque. Eccentricities shown are 0 (red, solid), 0.04 (green, dashed), 
0.08 (blue, dotted), 0.12 (magenta, dot-dash), 0.16 (cyan, solid), 
0.20 (black, dashed). Note the Lindblad torque reversal for $e > 1.1h$ manifested as 
the reversal of the sign of the torque contributions from both the inner and outer disc. 
Also note the clear localisation of the corotation torque to the corotation region.}
 \label{fig::torque_density}
\end{figure}

For the purpose of calculating the corotation torque using method (ii),
the corotation region is taken to be an annulus whose width is defined to be the distance 
between the points on the inner and outer last circulating streamlines that are furthest from the planet's orbital radius.
Once the corotation region has been defined, the gravitational force exerted by disc material
on the planet from within that region is summed and time-averaged.
Given that we are interested in measuring
the steady corotation torque we apply this method to simulations in set B only.
In general we expect the corotation
torques to be localised within this horseshoe region, and the Lindblad torque to originate 
from beyond a distance to the planet equal to $2H/3$. In figure \ref{fig::torque_density}, 
we show the torque acting on the planet as a function of radius in the disc, 
demonstrating the localisation of the two torque contributions to these regions.
Further features displayed in the figure are worthy of note. We see the magnitude
of the corotation torque decrease with increasing eccentricity, and we also observe
the Lindblad torque contributions from the inner and outer disc change
sign as the eccentricity exceeds $h$. Furthermore, for the higher eccentricity
cases we see the contributions from the inner and outer disc torques originate
from the pericentre and apocentre of the planet orbit.

\begin{figure}
 \includegraphics[width=0.5\textwidth]{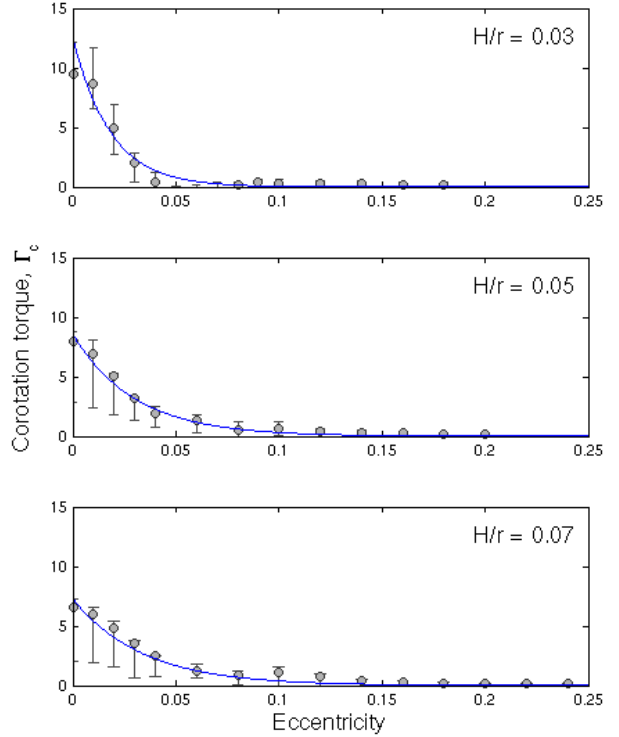}
 \caption{We plot the corotation torque measured by method (ii) by summing the steady-state torque applied 
to the planet from the material inside last circulating streamlines. Calculations are performed for 
simulation set B, with a sustained corotation torque. Error-bars are calculated by performing the 
same calculation with the boundaries moved 25\% further away from the planet and 25\% closer 
respectively. Note that this method underestimates the torque compared to the other two. Results 
are given for three disc aspect ratios and fits are of the form $\Gamma_c = \Gamma_{c,e=0}
\exp\left(-e/e_{\rm f}\right)$. 
Corotation torque is given in units of $\Gamma_0/\gamma$.}
 \label{fig::method_ii_and_fit}
\end{figure}

The estimates of the steady corotation torques for each of the models with planet 
mass equal to 5 $\Me$ and $h=0.03$, 0.05 and 0.07 are shown in
figure \ref{fig::method_ii_and_fit}. The filled circles represent values for
the corotation torque obtained by taking a fiducial value for the width of
the horseshoe region (the distance 
between the points on the inner and outer last circulating streamlines that are furthest from the planet's orbital radius). The error bars represent the 
fact that there is some ambiguity in the corotation torque because the last circulating 
streamlines used to define the boundary of the corotation region do not 
lie at constant distance from the corotation radius of the planet guiding centre. 
These error bars were obtained by 
moving the boundary of the corotation region 25\% further away from the planet 
and 25\% closer to it. 

As with method (i) for estimating corotation torques, this method also suffers 
from a drawback, which is that high density material that forms close to the
planet at apo- and pericentre can enter the defined horseshoe region. Even 
though these high density features are not related to the horseshoe drag, they
nonetheless can contribute to the estimate of the torque using method (ii) because
we have no way of excluding them from the torque calculation. In terms of the
magnitude of the corotation torque estimate, this method gives a lower value
than the other two because of this effect.
We note, however, that this method gives a smoothly varying monotonic estimate of
the corotation torque as a function of eccentricity, unlike methods (i) and (iii), 
demonstrating that the steady corotation torque in a viscous disc with cooling really does
behave in the expected manner.

\subsection{Method (iii): Initial peak of torque time series}

\begin{figure}
 \includegraphics[width=0.5\textwidth]{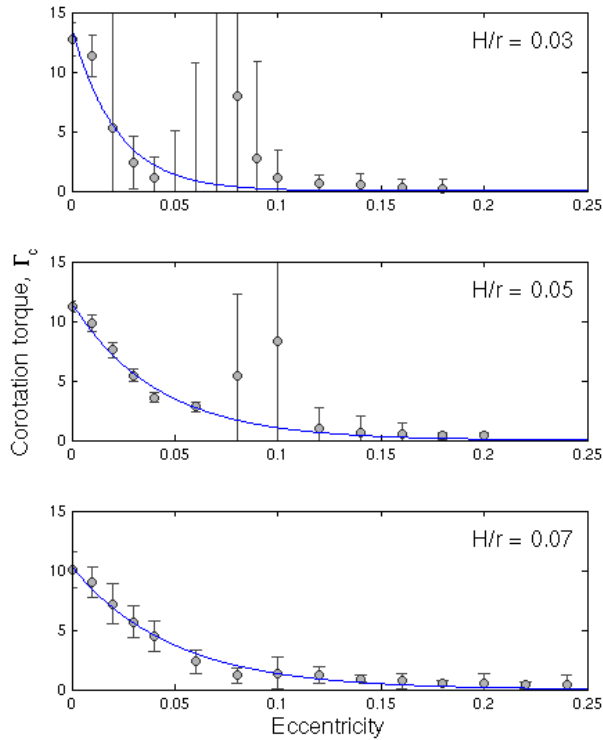}
 \caption{We plot the corotation torque measured by method (iii) described in the text.
Error-bars are three times the combined standard deviation of the initial peak and the 
steady state value. Results are given for three disc aspect ratios $H/r=0.03$, 0.05 and 0.07,
 and fits are of the form $\Gamma_c = \Gamma_{c,e=0} \exp\left(-e/e_{\rm f}\right)$. Corotation torque is given 
in units of $\Gamma_0/\gamma$.}
 \label{fig::method_iii_and_fit}
\end{figure}

This method is the most comparable to that used in previous work 
\citep[e.g.][]{paardekooper2009torque}. We begin by using a Fourier 
transform filter to remove high frequency oscillations from torque time series obtained
from simulations in set A. We then measure the difference between the long-term steady 
state (Lindblad) torque in these low viscosity adiabatic discs, and the torque value immediately 
after approximately one horseshoe libration period has elapsed when the surface density 
perturbations in the horseshoe region have been set up through the advection of entropy and vortensity. 
This is the moment when the transient corotation torque reaches its maximum positive value,
as shown for example by the dashed line in figure~1. Corotation torque estimates obtained
using this method are shown in figure \ref{fig::method_iii_and_fit}.
As with method (i), this method also has some drawbacks, because the long-term torque
that is supposed to represent the Lindblad torque is influenced by the previously described
gap and vortex formation.

\subsection{A thicker disc: $h=0.1$}

\begin{figure}
 \includegraphics[width=0.5\textwidth]{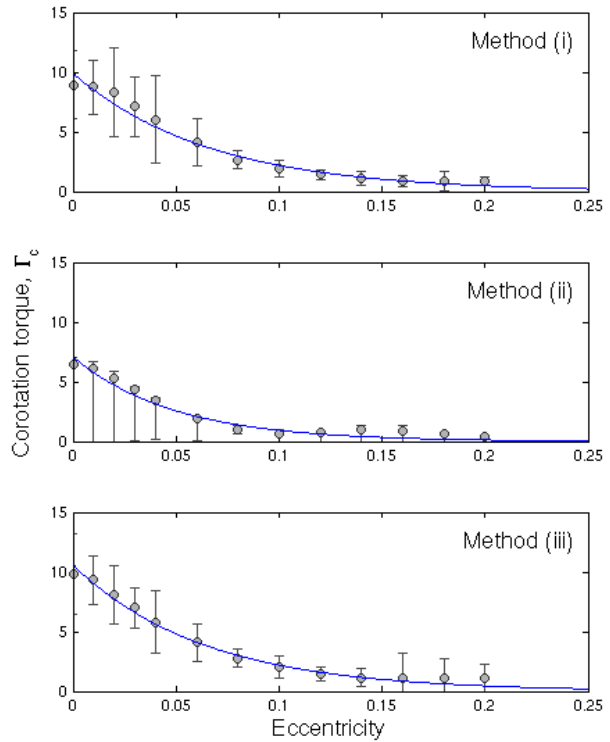}
 \caption{For the disc with the largest aspect ratio, $h=0.1$, we calculate the corotation torque 
using the three methods detailed in the test. The fits are of the form $\Gamma_c = \Gamma_{c,e=0}\exp\left(-e/e_{\rm f}\right)$. 
Corotation torque is given in units of $\Gamma_0/\gamma$.}
 \label{fig::h010_torques}
\end{figure}

The simulations described above adopted discs with aspect ratios in the range
expected for protoplanetary discs. Our results, however, show significant dependence
on the disc thickness because of non-linear effects, so we consider a thicker disc model
with $h=0.1$. We have repeated the corotation torque estimates obtained from
methods (i), (ii) and (iii) for a broad range of eccentricities, and the corotation
torque values are plotted in figure \ref{fig::h010_torques}. The plots in this figure
confirm the general trends noted for the thinner disc models: improvement in the
behaviour of torque estimates as one employs thicker disc models; a tendency for
method (ii) to produce a low estimate for the corotation torque; and clear
decrease in corotation torque as the eccentricity increases.

\section{Discussion}

In this section, we analyse the fits that predict exponential decay of the corotation torque
with increasing eccentricity that we have shown in the figures presented in the previous section,
and discuss discrepancies between some of our simulation results and this trend. We 
also discuss some limitations of our experimental method, and go on to show how our results 
are broadly consistent with previous work relating the corotation torque to the horseshoe width.

\subsection{Fitting formulae}
Before discussing the fitting procedure, we recall that the total corotation 
torque is given as a sum of the barotropic and entropy-related contributions:
$\Gamma_{\rm c}=\Gamma_{\rm c,baro} + \Gamma_{\rm c,ent}$ \citep{paardekooper2011torque}.
Furthermore, these contributions to the unsaturated horseshoe drag scale with the 
width of the horseshoe region according to $\Gamma_{\rm c} \sim x_{\rm s}^4$.
Both $\Gamma_{\rm c,baro}$ and $\Gamma_{\rm c,ent}$ depend on the relative time scales
associated with horseshoe libration and the viscous/thermal diffusion time scales,
as these determine the level of torque saturation. As we have discussed already,
the width of the horseshoe region, $x_{\rm s}$, depends on the planetary eccentricity,
so we might expect the magnitude of the corotation torque for %
an eccentric planet in a disc with fixed thermal and viscous evolution times 
to decrease through the $x_{\rm s}^4$ dependence, and to also decrease compared to
the circular orbit case through changes in the level of torque saturation.

In principle it is possible to disentangle these two effects when fitting the
results of the simulations, but this would require a CPU-intensive programme
of runs in which the optimal values for the viscosity and thermal relaxation are
sought for each value of planet eccentricity. We avoid this complication by
fitting a simple function to the simulation results.

\begin{figure}
 \includegraphics[width=0.5\textwidth]{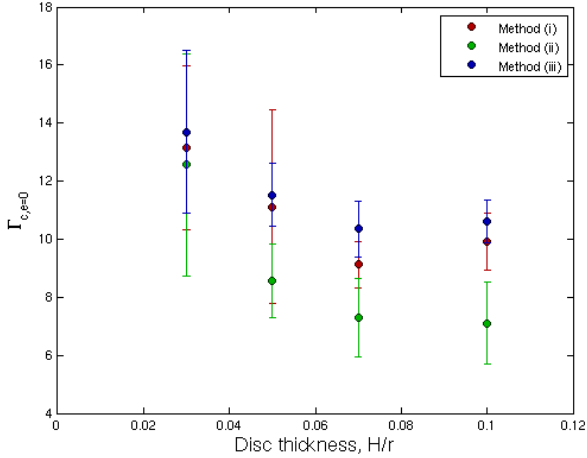}
 \caption{For the twelve simulations plotted as figures 4, 8, 9 and 10, we take the fits of the 
form $\Gamma_{\rm c} (e) = \Gamma_{\rm c,e=0} \exp\left(-\frac{e}{e_{\rm f}}\right)$ and plot $\Gamma_{\rm c,e=0}$
as a function of disc aspect ratio. Error-bars are taken from the 95\% confidence intervals of the fits.}
 \label{fig::a_fit}
\end{figure}

Denoting the corotation torque for a zero-eccentricity orbit as $\Gamma_{c,e=0}$,
normalised by $\Gamma_0/\gamma$, where we remind the reader that $\Gamma_0$ is given by equation \ref{eq:gam0},
we fit the torque as a function of eccentricity using the expression
\begin{equation}
\Gamma_{\rm c}(e) = \Gamma_{c,e=0} \exp\left(-\frac{e}{e_{\rm f}}\right).
\label{eq:gamma}
\end{equation}
We note that because of the normalisation by $\Gamma_0$, the zero-eccentricity corotation torque
is expected to be independent of $q$ and $h$ when in the linear regime. 
This is because the horseshoe width is expected to scale as $x_{\rm s} \sim \sqrt{q/h}$,
cancelling the $(q/h)^2$ dependence contained in $\Gamma_0$.
For larger values of $q$, or small values of $h$, however, the width of the corotation torque 
increases because of its sensitivity to the relative strengths of planet gravity and thermal 
pressure \citep{2006ApJ...652..730M,paardekooper2009width}. This causes $\Gamma_{\rm c,e=0}$
to increase in our simulations as $h$ decreases, as shown in figure~\ref{fig::a_fit}
where we plot $\Gamma_{\rm c,e=0}$ versus $h$.
We note that for $h=0.05$, our value of  $\Gamma_{\rm c,e=0}$ obtained with methods (i) and 
(iii) agrees well with the canonical value of 11.25 shown in figure~17 of \citet{paardekooper2009torque},
who also examined this issue. Furthermore, in that paper it was suggested that as $q$ increases
or $h$ decreases, it is appropriate to change the value of the coefficient $C$ in the expression
\begin{equation}
\label{eq:xs}
\frac{x_s}{r_{\rm p}} =C \left(\frac{b/h}{0.4}\right)^{1/4} \sqrt{\frac{q}{h}}
\end{equation}
from $C=1.1$ in the fully linear regime to 1.3 in the quasi-nonlinear regime to account for
this effect in the torque formulae.

\begin{figure}
 \includegraphics[width=0.5\textwidth]{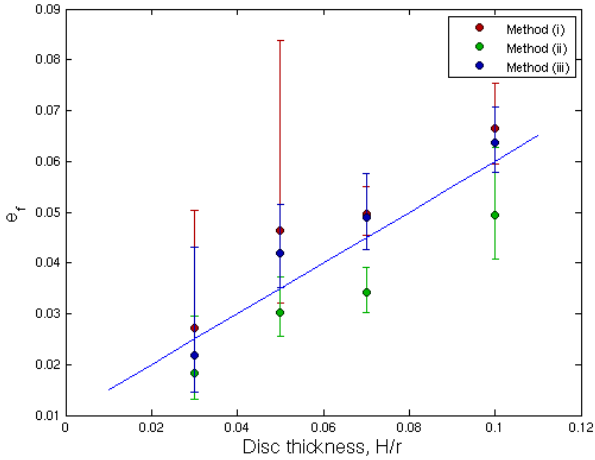}
 \caption{For the twelve simulations plotted as figures 4, 8, 9, 10, we take the fits of the form 
$\Gamma_{\rm c} (e) = \Gamma_{\rm c,e=0} \exp\left(-\frac{e}{e_{\rm f}}\right)$ and plot $e_{\rm f}$ as a 
function of disc aspect ratio. Error-bars are taken from the 95\% confidence intervals of the fits. 
We superimpose a simple linear fit of $e_{\rm f} = h/2 + 0.01$.}
 \label{fig::b_fit_line}
\end{figure}

We now consider fitting the characteristic e-folding eccentricity, $e_{\rm f}$.
In figure \ref{fig::b_fit_line} we plot the best fitting values of $e_{\rm f}$
obtained using each of the methods (i), (ii) and (iii) as a function of the disc aspect
ratio $h$. Methods (i) and (iii) give very similar values, and method (ii) gives
values that are offset slightly but with a similar slope. All methods give an 
approximately linear relation between $e_{\rm f}$ and $h$. The superimposed line
in the plot is given by
\begin{equation}
e_{\rm f} = h/2 + 0.01,
\label{eq::char_eccen}
\end{equation}
Using this relationship, and the one in equation \ref{eq:gamma}, it is possible to 
obtain the corotation torque attenuation experienced by an eccentric planet.

\subsection{A higher mass planet}
\begin{figure}
 \includegraphics[width=0.5\textwidth]{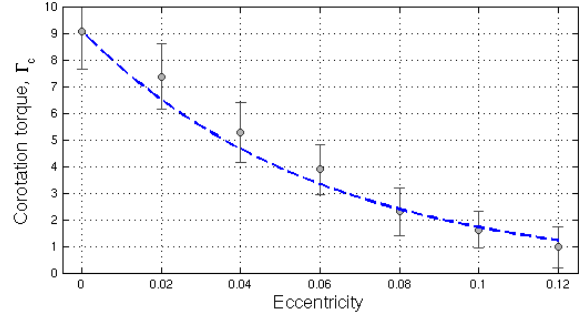}
 \caption{Corotation torque for a $10 \, \Me$ planet embedded in a $h = 0.10$ disc, measuring using 
our method (iii). Error bars are given to 1$\sigma$. We superimpose a fit of the form 
$\Gamma_{\rm c}(e) = \Gamma_{\rm c,e=0} \exp\left(-e/e_{\rm f}\right)$, using $e_{\rm f}$ from equation 
\ref{eq::char_eccen}, derived from a study of $5\, \Me$ planets.}
 \label{fig::10Me}
\end{figure}

So far we have only considered variations in the disc aspect ratio
and planet eccentricity, for which the fitting formulae presented 
in the previous section provide good overall fits to the data,
as shown in the figures \ref{fig::method_i_and_fit}, \ref{fig::method_ii_and_fit}, \ref{fig::method_iii_and_fit}
and 10. We now demonstrate that these fits also give good results when applied to a planet
with 10 $\Me$ instead of $5 \, \Me$. Figure~\ref{fig::10Me} shows
the corotation torque estimated using method (iii) from simulations with 
$h=0.1$ and a planet with $m_{\rm p} = 10 \Me$. We observe that the fitting
formulae given by equations \ref{eq:gamma} and \ref{eq::char_eccen}
give very satisfactory results for this case, suggesting that
they can be used for a broad range of planet masses, eccentricities
and disc aspect ratios. In particular, we note that the
characteristic e-folding eccentricity depends only on the disc aspect 
ratio and not on planet mass, at least for the range of parameters
that we have considered.

\begin{figure*}
\begin{minipage}{140mm}
 \includegraphics[width=\columnwidth]{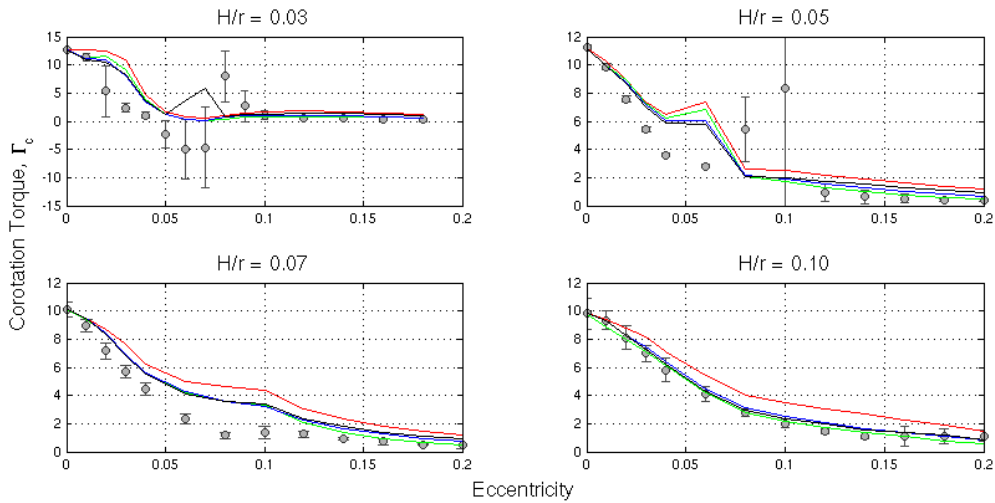}
 \caption{For all our disc aspect ratios we plot corotation torque against eccentricity 
in units of $\Gamma0/\gamma$, as measured by method (iii). Error bars are now only one 
standard deviation. We superimpose lines representing different measures of the horseshoe 
width. The red line is the width measured at azimuth, $\theta=0$; the green line 
corresponds to $\theta=3\pi/2$; the blue line refers to the horseshoe width for $\theta$ 
just beyond the location of the dip due to the planet's location; the black line is the 
maximum value of the horseshoe width.}
 \label{fig::all_scaledhshoe}
 \end{minipage}
\end{figure*}

\subsection{Physical interpretation}
Material at the edge of the horseshoe region, orbiting at a radial distance of $x_s$ from the
planet's location, will have a horseshoe libration time of
\begin{equation}
\tau_{lib} = \frac{8 \pi a_p}{3 x_s \Omega_p}.
\label{libtimescale}
\end{equation}
As this is much longer than the orbital period of the planet, and its epicyclic motion in the
rotating frame, we can say that material in the horseshoe region interacts with the planet
on time scales that are long compared to the orbital period. The planet's potential,
as experienced by the material librating with respect to it on horseshoe orbits,
may therefore appear softened due to its periodic radial excursion from the
corotation radius when averaged over one horseshoe libration period. We tentatively suggest
that this softening of the potential is responsible for the observed narrowing of the
horseshoe width, $x_s$, as the eccentricity increases. An alternative hypothesis for the observed
narrowing of the horseshoe region is that the excursion in azimuth of the planet as it
undergoes epicyclic motion causes the horseshoe streamlines that approach the planet most closely
to be disrupted. Given that these streamlines are the ones that define the outer edge of the
horseshoe region, this would cause the horseshoe region to narrow. We have examined the horseshoe
streamlines in some detail for increasing values of the planet eccentricity and can confirm
that this is not the case. Instead, we observe that as the eccentricity increases the
azimuthal location of the horseshoe u-turns moves away from the planet in a smooth manner.

The interpretation that the corotation torque decreases with increasing
eccentricity because of effective gravitational softening leads us to view
the eccentricity as the dimensionless length scale associated with epicyclic
motion. Consequently, we expect on physical grounds that the e-folding eccentricity,
$e_{\rm f}$, will depend on a characteristic length scale in the problem. In a real
three-dimensional disc there are only two natural length scales that may influence the
corotation torque, these being the horseshoe width for a circular orbit, $x_s$, and
the local pressure scale height, $h$. In earlier work,
\citet{hellarynelson2012} suggested that the decrease in the corotation torque with increasing
eccentricity observed by \citet{bitschkley2010} was due to the planet moving outside of
the horseshoe region, leading to the assumption that the important parameter in the
problem is $e/x_s$. The simulations of \citet{bitschkley2010} adopted parameters such that
$x_s \sim h$, so determining whether $e/x_s$ or $e/h$ is the important parameter
is difficult from their work. Our simulations have been designed to specifically address
this question, and show unambiguously that $e/h$ is the important parameter
because the e-folding eccentricity is a linear function of the scale height
through $e_{\rm f} = h/2+0.01$. Indeed, $x_s$ decreases as $h$ increases due to the pressure
in the disc acting as a buffer against the gravitational potential of the planet, so there
is no room for doubt from our simulations about whether it is $e/h$ or $e/x_s$ that
controls the rate at which the corotation torque decreases as the eccentricity increases. 
The buffering influence of the pressure
explains why $e/h$ determines the magnitude of the corotation torque: the eccentricity
of the planet is competing with the pressure in determining the width of the corotation
region, so for appreciable changes in $x_s$ to occur it seems that $e$ must be 
comparable to $h$.

Our simulations are two-dimensional and require the use of a gravitational softening
parameter, $b$, whose primary role is to allow two-dimensional results to agree with
three-dimensional simulations by accounting for missing 3D effects. Normally,
$b$ is chosen to be a linear function of $H$, with values typically being on the order of
$0.4 H$ as in this work. The introduction of $b$ brings another length scale into the
problem that may influence the scaling of the corotation torque with eccentricity.
We present a suite of runs in appendix A to examine this, where the scale height remains
constant at $h=0.07$, $b/h$ takes values from 0.2 up to 0.8, and for each value of
$b/h$ the eccentricity takes on values between $e=0$ and $e=0.12$. We follow the same
procedure described in section 4.1 in obtaining a fit to the corotation torque
($\Gamma_{\rm c}(e) = \Gamma_{c,e=0} \exp{\left[-e /e_{\rm f}\right]}$), and examine
whether or not $e_{\rm f}$ can be expressed as a linear function of dimensionless $b$. 
Our results demonstrate that this is not the case. At best 
$e_{\rm f}$ is a very weak function of $b$, and is consistent with our
original fit $e_{\rm f}=h/2+0.01$. 
This result demonstrates that it is the scale height, $h$, and not the softening, $b$, 
that determines the behaviour of the corotation torque as $e$ increases in our simulations.
Although the reason for this is not entirely clear, we suggest that the primary reason is 
that the width of the horseshoe region $x_s$ is being controlled primarily by the scale height,
$h$, rather than the softening parameter, $b$, in most of our runs, so that the softening effect 
introduced by increasing the eccentricity is competing with $h$ rather than $b$.
Some support for this interpretation is provided by the fact that $x_s$ has a stronger
functional dependence on the scale height, $h$, than on the softening, $b$.

If this interpretation is true then it implies that there is a range of values of $b$ for
which the softening plays the most important role in controlling $x_s$, and for that range of 
values we would expect $e/b$ to control the rate at which the corotation torque decreases as 
$e$ increases. The values of $b$ for which this is true are likely to be significantly larger
that $0.4 H$, meaning that this parameter regime lies outside of the range of models that
closely mimic the behaviour expected for 3D simulations that require $b \sim 0.4 H$. 

We note that the previously mentioned gravitational softening due to the planet's epicyclic 
motion can be observed to operate in our simulations by comparing runs on a case by case basis.
For an eccentric orbit the apparent softening length is $b_*^2 = b^2 + e^2a^2$.
We therefore expect that, for example, a $5\Me$ planet on a circular orbit in a disc with $h=0.07$
with $b/h=0.8$ will exhibit the same corotation torque as a run with $b/h=0.4$ and $e=0.0485$,
and this is indeed found to be the case in our runs within the margin of error involved
in measuring corotation torques. This adds further weight to the physical interpretation described 
above.

Finally, we now discuss how consistent our results are with the interpretation that
the corotation torque decreases because the horseshoe width narrows with increasing
eccentricity. As mentioned already in the analysis by Ward (1991), and later work by
\citet{2001ApJ...558..453M} and \citet{paardekooper2009torque}, the width of
the corotation region is related to the corotation torque by the scaling:
\begin{equation}
\Gamma_c \sim x_s^4.
\end{equation}
In figure \ref{fig::all_scaledhshoe}, we attempt to fit just such a scaling to our torque
measurements obtained using method (iii), normalised to match the measured torque at
zero-eccentricity.
Although the overall scaling, covering the full range of eccentricity values considered,
is reasonably well captured by the curves, it is clear that the corotation torques
in the simulations fall off faster than predicted by the $x_s^4$ scaling.
One possible explanation for this is that the narrowing of the horseshoe
region causes the thermal relaxation time and viscosity in the simulations
to be no longer optimal for unsaturating the torque, leading to a further
reduction in its value beyond the fall off predicted by the $x_s^4$ scaling.

\subsection{Corotation torque set-up timescale}
A planet on a circular orbit migrating because of tidal interaction with the disc will 
retain the material in the horseshoe region as it migrates. Consequently the corotation 
torque will evolve gradually as the semi-major axis changes.
A planet that experiences a very rapid change in its position in the disc, however,
due to planet-planet scattering, will set up a new corotation region with material
undergoing horseshoe libration. The scattering will likely leave the planet in
an eccentric orbit initially when it lands at the new semi-major axis, so the growth of the
new corotation torque will occur on the time scale for eccentricity damping, followed
by the libration time scale given by equation \ref{libtimescale} as the planet
tends toward a circular orbit. In general, standard type I migration time scales are 
on the order of $10^4$ orbits for 1~$\Me$ planets, and $\sim 10^3$ orbits for 10 $\Me$ bodies
\citep[e.g.][]{2002ApJ...565.1257T}. The eccentricity damping time scale is typically
a factor $\sim (H/r)^2$ shorter than the migration time \citep{2004ApJ...602..388T}, 
bringing it close to the time scales for horseshoe libration for low mass planets.
As such, the damping of eccentricity and growth of the corotation torque will occur on
similar time scales. In principle, these are issues that should be accounted for
in N-body simulations of planetary formation that include prescriptions for
corotation torques, if planet-planet scattering plays an important role.

\section{Conclusion}
In this paper we have presented a suite of simulations that were designed to
examine how the steady disc-induced corotation torque varies as a function
of planet orbital eccentricity for low mass planets embedded in protoplanetary discs.
We considered disc models with four different
aspect ratios, and used three different methods to estimate the corotation torque.
In agreement with previous work \citep{bitschkley2010}, we find that the 
corotation torque decreases as the orbital eccentricity increases.
We provide an analytical fit to the numerically-obtained corotation 
torques as a function of eccentricity, and find that they are well-fitted by 
a simple exponential decay with e-folding eccentricity that scales linearly 
with the disc aspect ratio.

Through inspection of time-averaged fluid streamlines we find that the
fluid in the corotation region continues to undergo horseshoe orbits
when the eccentricity is non-zero. As the eccentricity increases we find
that the horseshoe region narrows, and we suggest that this is the major
reason why the corotation torque decreases with increasing eccentricity,
since the non-linear horseshoe drag, $\Gamma_c$, is known to scale as 
$\Gamma_c \sim x_s^4$. When plotting the measured values of $\Gamma_c$
against the measured values of $x_s$ we find that the corotation
torques from the simulations drop off more rapidly than suggested by
the $x_s^4$ scaling. We tentatively suggest that the changing width of
the horseshoe region causes the adopted values of thermal relaxation time 
and viscosity in the simulations to become suboptimal for fully unsaturating 
the corotation torque, causing the torque to be smaller than predicted by
the $x_s^4$ scaling.

While previous work \citep[e.g.][]{hellarynelson2012} has made use of a simple model of 
corotation torque as a function of eccentricity, wherein the parameter governing the torque 
attenuation is $e/x_s$, we have shown instead that the torque decays 
as $e/e_{\rm f}$, where $e_{\rm f}$ can be modelled as a linear function of the disc
aspect ratio. This latter scaling produces a less severe drop-off in the magnitude of the
corotation torque with eccentricity, as the scale height is generally larger than the 
horseshoe width for low mass embedded planets. The fitting formula we have provided
should therefore provide a useful addition to N-body models of planet formation
that implement type I migration prescriptions including corotation torques,
especially if planet-planet scattering events are important.

Our results have implications for the notion of ``zero torque radii'' occurring in discs at 
locations where the (outward) corotation torque balances the (inward) Lindblad torque. 
Such locations may be important during planetary formation by acting as `traps' where
planetary building blocks may congregate, enhancing accretion. While the locations of these 
zero torque radii depend on the the properties of the local disc sufficiently optimising 
the corotation torque,  we have shown that a relatively modest planetary eccentricity can 
have an effect on the torque experienced by the planet, moving the location of zero torque 
radii, or even removing them entirely if the eccentricities become large enough, 
resulting in a qualitative effect on planetary migration and formation. One particular
scenario where this may be important is in the formation of circumbinary planets,
where the disturbing influence of the central binary may excite significant eccentricities,
as considered recently by \citet{2013arXiv1307.0713P} in application to the Kepler-16, 34 and 35
systems.

Furthermore, as eccentricities are often excited by planetary bodies in mean motion resonance, 
our work has implications for pairs of planets being able to remain in resonance after having 
their eccentricities excited, and therefore on the subsequent evolution of such a system. 
For example, a pair of planets may migrate convergently into resonance, because of 
the influence of corotation torques, excite their mutual eccentricities,
and then migrate divergently such that the resonance is not maintained.
Subsequent damping of the eccentricity will then cause this process to repeat,
keeping the system near to, but not actually in resonance. Such a mode of evolution
could potentially explain the compact systems of low-mass planets discovered
by the Kepler mission (e.g. Kepler-11 \citep[][]{2011Natur.470...53L}) which are
close to, but not in resonance.

This work has been limited in using a simple thermal model in a 2D disc.
In future work we plan to revisit some of the issues raised in the paper
using 3D models of discs with radiative transfer.

\vspace{-6mm}
\section*{Acknowledgements}

SMF acknowledges the support of an STFC PhD studentship. The simulations 
presented in this paper were performed on the QMUL HPC facility purchased
under the SRIF initiatives.

\vspace{-6mm}
\bibliographystyle{mn2e}
\bibliography{mn2epaper}

\onecolumn
\appendix
\section{Dependence on the softening parameter, \lowercase{$b/h$}}
\label{subsect:h}

In order for 2D hydrodynamic simulations to produce results comparable to 3D ones, 
it is necessary to introduce a gravitational softening parameter, $b$, to compensate 
for absent 3D effects. This parameter is taken to be a linear function of the local height
of the disc, typically on the order of $0.4h$ as in this paper. 

\begin{figure*}
 \includegraphics[width=0.32\textwidth]{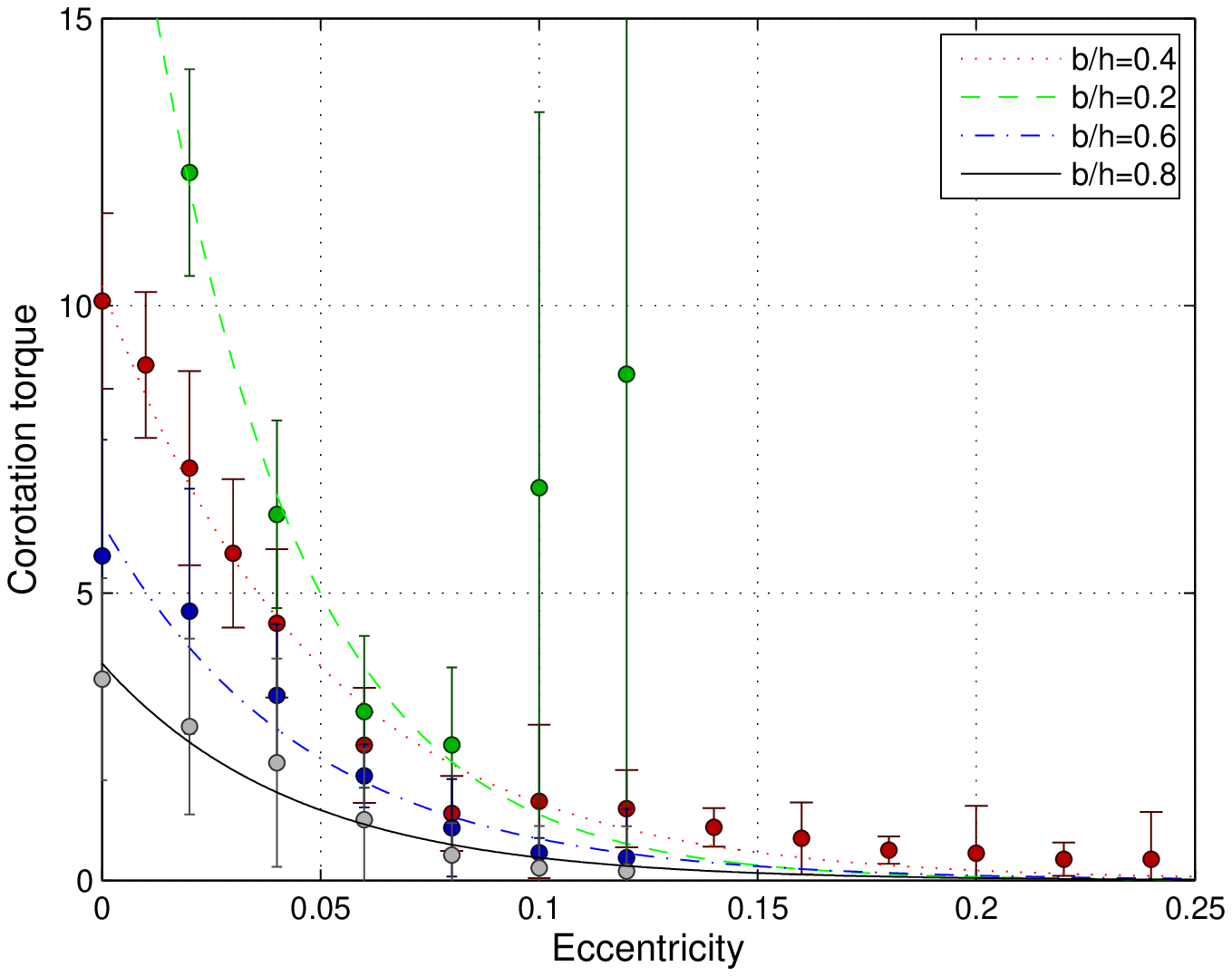}
\includegraphics[width=0.32\textwidth]{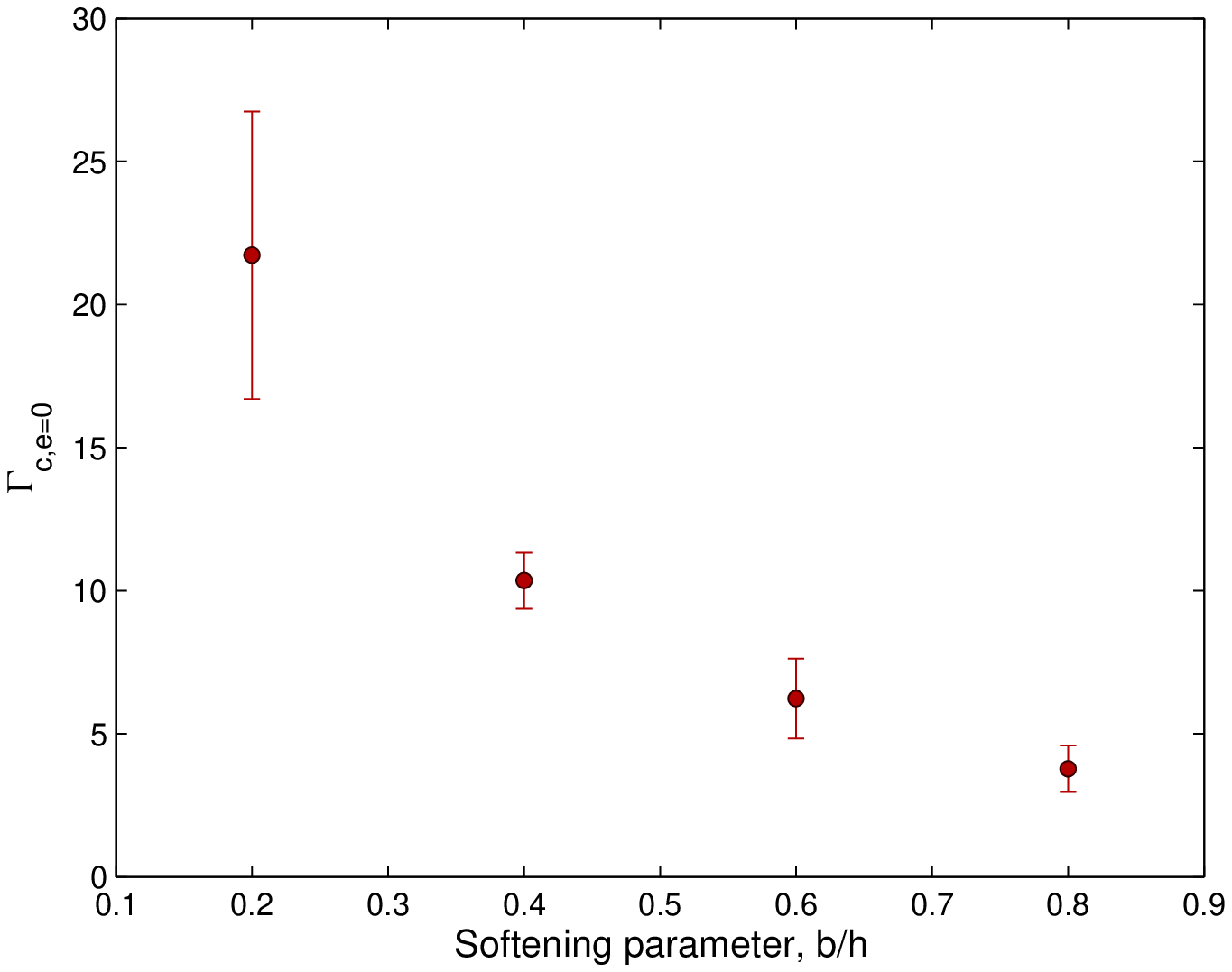}
\includegraphics[width=0.32\textwidth]{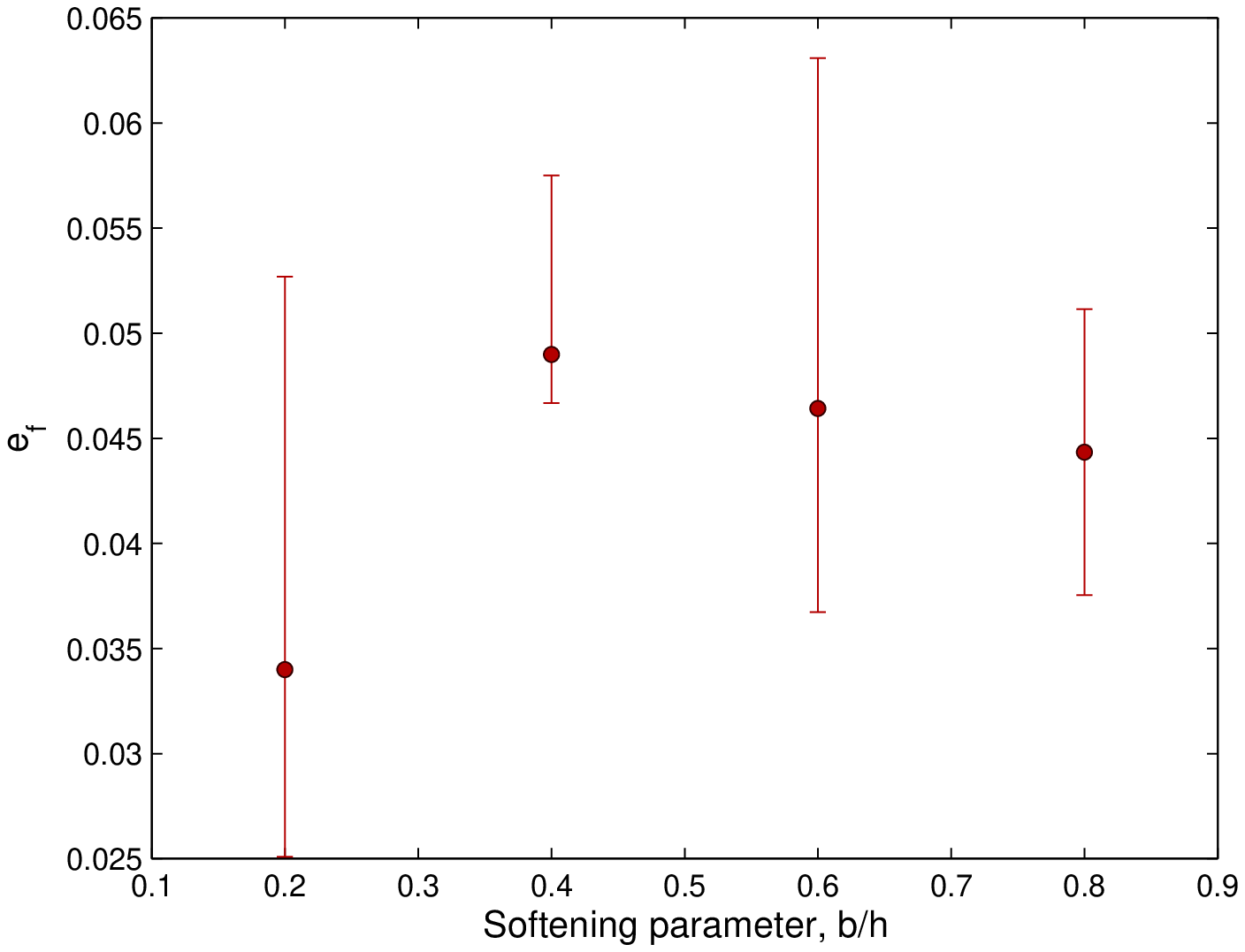}
 \caption{Left panel: the corotation torque on a 5$\Me$ planet in units of 
$\Gamma_0/\gamma$ obtained using method (iii). We superimpose fits of the form 
$\Gamma_c = \Gamma_{c,e=0}\exp\left(-e/e_{\rm f}\right)$ over these data.
Middle panel: We show the parameter $\Gamma_{c,e=0}$ from the fits shown in the left panel, 
as a function of softening, $b/h$.
Right panel: We show the parameter $e_{\rm f}$ from the fits shown in the left panel, 
as a function of the softening, $b/h$.}
 \label{fig::apdx_fits}
\end{figure*}

Given  our tentative physical explanation of why the corotation torque decreases with eccentricity given
in Sect.~4.3, based on the idea that the epicyclic motion of the planet induces an effective softening of
the planet potential, here we examine whether or not the softening parameter, $b$, or the scale height, $h$,
are most important for setting the scaling of $e_{\rm f}$, the e-folding eccentricity used in our
analytical fits. To this end, we ran a series of simulations of a 5$\Me$ planet embedded in 
a $h=0.07$ disc, with values of $b/h$ between 0.2 and 0.8. We again perform fits of the form 
$\Gamma_c = \Gamma_{c,e=0}\exp\left(-e/e_{\rm f}\right)$,
and show these superimposed on the data in the left panel of figure~\ref{fig::apdx_fits}. 
We note that using a small value of $b/h=0.2$ can also lead to non-linear restructuring of the disk, 
similar to described in Sect.~3.1, explaining the outlying points at eccentricity values 
$e=0.1$ and 0.12.

The values we obtain for the parameter $\Gamma_{c,e=0}$ are shown in the middle 
panel of figure~\ref{fig::apdx_fits}. We note that in the circular orbit case, we 
expect the corotation torque to scale as $x_s^4$, and therefore 
as $\frac{1}{b/h}$, and our data are consistent with this.
The e-folding eccentricity values, $e_{\rm f}$, are shown in the right panel
of figure~\ref{fig::apdx_fits} as a function of $b/h$.
We observe no obvious trend, and the best fitting values appear to be independent of $b/h$.
We note, however, that using equation ~10 ($e_{\rm f} = h/2 + 0.01$) yields a value $e_{\rm f}=0.045$,
which is consistent with the results shown in the right panel of figure~\ref{fig::apdx_fits}.

We conclude that the key physical quantity that determines the behaviour of $e_{\rm f}$ is 
the disc scale height $h$. The softening parameter, $b$, plays the important role of 
allowing 2D simulations to produce results that are consistent with 3D simulations, 
but does not play an important role in determining how the corotation 
torque scales with orbital eccentricity.

\bsp

\label{lastpage}

\end{document}